\documentclass[english,prl, superscriptaddress, twocolumn]{revtex4-1}
\usepackage[T1]{fontenc}
\usepackage[latin9]{inputenc}
\usepackage{babel}
\usepackage{amsmath}
\usepackage{amssymb}
\usepackage{graphicx}
\usepackage[bottom]{footmisc}

\makeatletter
\@ifundefined{textcolor}{}
{%
 \definecolor{BLACK}{gray}{0}
 \definecolor{WHITE}{gray}{1}
 \definecolor{RED}{rgb}{1,0,0}
 \definecolor{GREEN}{rgb}{0,1,0}
 \definecolor{BLUE}{rgb}{0,0,1}
 \definecolor{CYAN}{cmyk}{1,0,0,0}
 \definecolor{MAGENTA}{cmyk}{0,1,0,0}
 \definecolor{YELLOW}{cmyk}{0,0,1,0}
}

\makeatother

\begin{document}

\title{Quantum Noise Theory of Exceptional Point Amplifying Sensors}

\author{Mengzhen Zhang}

\affiliation{Departments of Applied Physics and Physics, Yale University, New
Haven, CT 06520, USA}

\affiliation{Yale Quantum Institute, Yale University, New Haven, CT 06520, USA}

\author{William Sweeney}

\affiliation{Departments of Applied Physics and Physics, Yale University, New
Haven, CT 06520, USA}

\affiliation{Yale Quantum Institute, Yale University, New Haven, CT 06520, USA}

\author{Chia Wei Hsu}

\affiliation{Departments of Applied Physics and Physics, Yale University, New
Haven, CT 06520, USA}

\affiliation{Yale Quantum Institute, Yale University, New Haven, CT 06520, USA}

\author{Lan Yang}

\affiliation{Department of Electrical and Systems Engineering, Washington University,
St Louis, Missouri 63130, USA}

\author{A. D. Stone}

\affiliation{Departments of Applied Physics and Physics, Yale University, New
Haven, CT 06520, USA}

\affiliation{Yale Quantum Institute, Yale University, New Haven, CT 06520, USA}

\author{Liang Jiang{*}}

\affiliation{Departments of Applied Physics and Physics, Yale University, New
Haven, CT 06520, USA}

\affiliation{Yale Quantum Institute, Yale University, New Haven, CT 06520, USA}
\begin{abstract}
Open quantum systems can have
exceptional points (EPs), degeneracies where both eigenvalues and eigenvectors coalesce.
Recently, it has been proposed and demonstrated that EPs can enhance
the performance of sensors in terms of amplification of a detected signal.
However, typically amplification of signals also increases the system noise, and 
it has not yet been shown that an EP sensor can have improved signal
to noise performance.  We develop a quantum
noise theory to calculate the signal-to-noise performance of an EP
sensor.  We use the quantum Fisher information to extract a lower bound
for the signal-to-noise ratio(SNR) and show that parametrically improved SNR
is possible.   Finally, we construct a specific 
experimental protocol for sensing using an EP amplifier near its lasing
threshold and heterodyne signal detection to achieves the optimal 
scaling predicted by the Fisher bound.  Our results can be generalized
to higher order EPs for any bosonic non-Hermitian system with linear interactions.
\end{abstract}
\maketitle

A distinct feature of physical systems described by non-Hermitian operators is the possibility of
exceptional points (EPs), degeneracies at which not only eigenvalues  
but also the eigenvectors coalesce \cite{kato_perturbation_2013,heiss_phases_1999,heiss_repulsion_2000,berry_physics_2004}.
EPs have been extensively discussed in the context of generalizing
the standard quantum theory to include non-Hermitian Hamiltonians
\cite{moiseyev_non-hermitian_2011}.   However we already can study the effect of EPs
in {\it open} systems within the conventional quantum framework, either by looking at resonant
scattering or by treating unobserved degrees of freedom as lossy or amplifying reservoirs.
Many interesting effects of EPs have been studied in electromagnetic or optomechanical systems \cite{lin_unidirectional_2011, chang2014parity, peng_paritytime-symmetric_2014,xu_topological_2016,doppler_dynamically_2016,makris_beam_2008,zhen_spawning_2015};
often these systems have parity-time symmetry\cite{ruter_observation_2010, regensburger_paritytime_2012, chang2014parity, feng_single-mode_2014, zhang_observation_2016}, although this symmetry is not essential
to the existence of EPs.

One intriguing application of EPs is to enhance the performance of
sensors, which has been theoretically proposed \cite{wiersig_enhancing_2014,wiersig_sensors_2016,liu_metrology_2016},
and experimentally demonstrated using optical micro-ring resonators
\cite{chen_exceptional_2017,hodaei_enhanced_2017}. Here we will focus
on the most important case of resonant detectors, for which the relevant 
non-hermitian operator is the open system Hamiltonian (cavity wave operator), and
the eigenvalues are the complex resonant frequencies of this operator.
Assume that the resonances are non-degenerate initially: by tuning several parameters of this operator, 
two or more resonances can be brought to degeneracy without imposing any additional symmetries, thus
creating an EP corresponding to only a single resonant solution at that complex frequency.  
One consequence of operating at a second-order EP is that a small perturbation,$\sim \epsilon$, which lifts the 
degeneracy, will split the resonant frequencies by $\sim \sqrt{\epsilon}$, a parametrically larger 
sensitivity than for hermitian systems.  This type enhanced frequency sensitivity was demonstrated
in the micro-ring experiments \cite{chen_exceptional_2017,hodaei_enhanced_2017}, but 
at present there has been no systematic analysis of the effect on the noise and the signal-to-noise
ratio (SNR) of operating a sensor in the vicinity of an EP.  According to quantum noise theory \cite{gardiner_quantum_2004},
the gain (and loss) introduced via non-Hermitian dynamics will unavoidably generate
additional noise, making it unclear that the EP sensor will have enhanced SNR (assuming 
that the sensor is operating near the quantum noise limit, and is not dominated by extrinsic noise sources). 
Therefore, in evaluating the efficacy of EPs sensors, it is crucial to calculate the behavior of both signal and
noise near an EP.

In this Letter, we present such an investigation, which addresses the following three
questions: (1) Can operating near an EP enhance the SNR of a sensor? (2) What
is the maximal precision (in terms of SNR) of EP sensing schemes? (3) How can we design
an EP sensing scheme to achieve this ultimate precision? To answer
these questions, we first apply quantum noise theory \cite{gardiner_quantum_2004}
to calculate the amplitude and covariance matrix associated with the
output of a model of EP sensors; then we calculate the quantum Fisher information
of the output state and obtain the Cramer--Rao bound for the parameter
estimation. Finally we explicitly construct an EP 
sensing scheme based on heterodyne detection of the output of an amplifying EP sensor to achieve this optimal scaling
of the sensitivity.  Our scheme is not set up to detect the
frequency splitting near an EP, but rather the amplitude change of a signal amplified near the lasing frequency of our
sensor.  The results can be generalized to higher order EPs
for any bosonic non-Hermitian system with linear
interactions, i.e. involving only Gaussian processes \cite{weedbrook_gaussian_2012}.

\begin{figure}
\includegraphics[width=1\columnwidth]{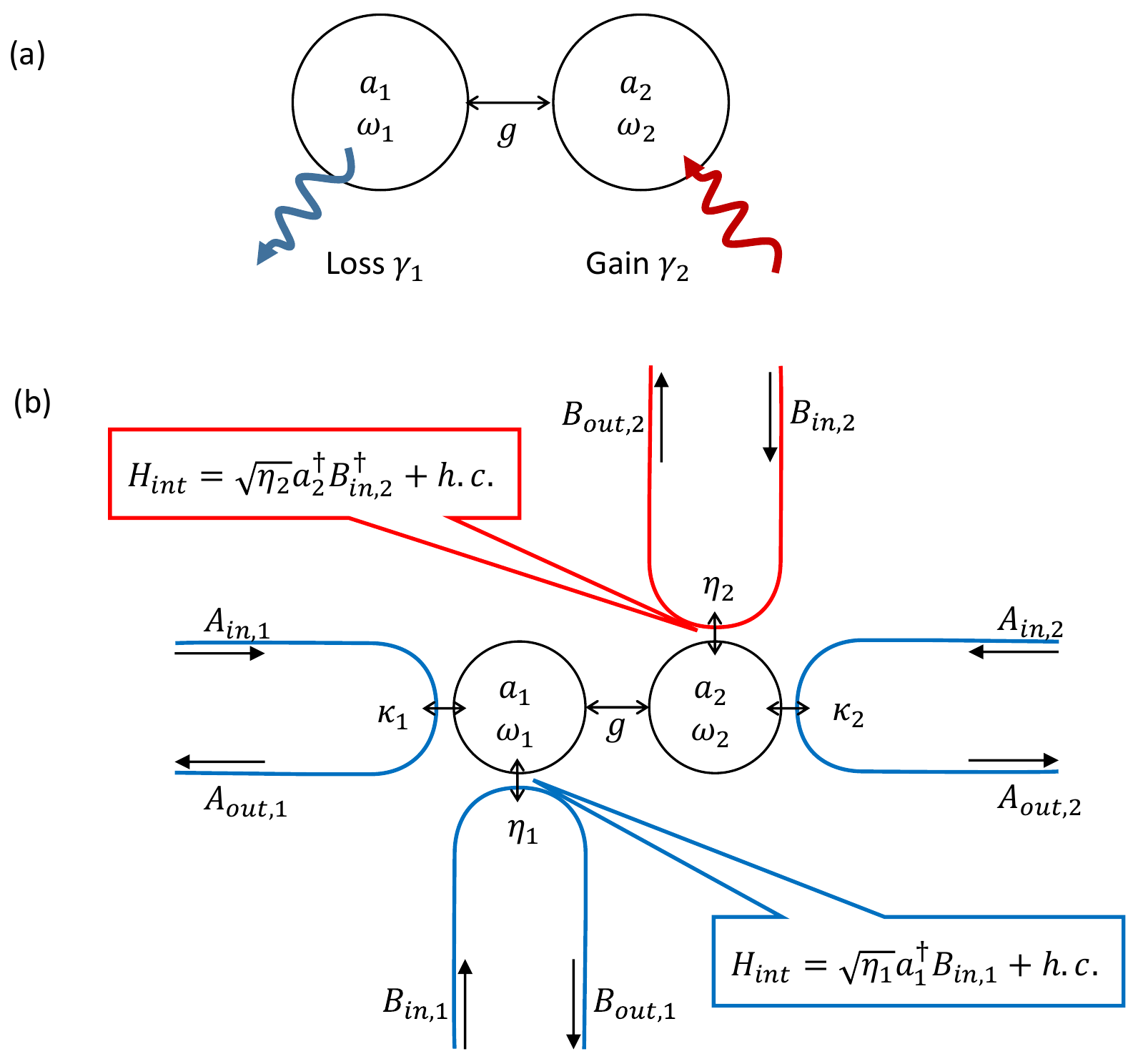}
\caption{\label{fig:1}\textbf{(a).} Schematic of two-bosonic-mode system with loss and gain
modeled classically. \textbf{(b).} Schematic of two-bosonic-mode system
with a full quantum description. The circles labeled represent two
bosonic modes $a_{1}$ and $a_{2}$, coupled to probing channels $A_{1}$
and $A_{2}$. In addition to the probing channels, $a_{1}$ is intrinsically
dissipated by scattering channel $B_{1}$ and $a_{2}$ is intrinsically
amplified by scattering channel $B_{2}$, with different coupling
interactions. The arrows in the channels which point towards the system
modes represent the input modes, while those with opposite direction
represent the output modes.}
\end{figure}

\textit{Non-Hermitian dynamics and open quantum systems.}
We define a non-Hermitian Hamiltonian to
characterize the open sensor \cite{el2018non}. A generic model is
shown in Fig.~\ref{fig:1}a; we have coupled cavities with two resonant optical modes
$a_{1}$ and $a_{2}$,  with the non-Hermitian Hamiltonian (setting
$\hbar=1$)
\begin{align}
\hat{H} = &\left(\omega_{1}+\epsilon_{1}\right)\hat{a}_{1}^{\dagger}\hat{a}_{1}+\left(\omega_{2}+\epsilon_{2}\right)\hat{a}_{2}^{\dagger}\hat{a}_{2}+g\left(\hat{a}_{1}^{\dagger}\hat{a}_{2}+h.c.\right)\nonumber \\
 & -i\frac{\gamma_{1}}{2}\hat{a}_{1}^{\dagger}\hat{a}_{1}+i\frac{\gamma_{2}}{2}\hat{a}_{2}^{\dagger}\hat{a}_{2}, 
\end{align}
where the total loss /gain rates of modes 1,2 are $\gamma_{1},\gamma_{2}$ and g is the inter-cavity
coupling.  To implement our EP sensing scheme we assume that
we can tune the real part of the resonance frequencies such that
$\omega_{1}=\omega_{2}=\omega$, and that the perturbation we are sensing
uniformly shifts the frequencies of both modes by $\epsilon_{1}=\epsilon_{2}=\epsilon$.  
The resulting equation of motion is
\begin{equation}
\frac{d}{dt}\begin{pmatrix}a_{1} \\
a_{2}
\end{pmatrix}  =  -i\begin{pmatrix}\left(\omega+\epsilon\right)-i\frac{\gamma_{1}}{2} & g\\
g & \left(\omega+\epsilon\right)+i\frac{\gamma_{2}}{2}
\end{pmatrix}\begin{pmatrix}a_{1}\\
a_{2}
\end{pmatrix} \label{eq:EOM-2}
\end{equation}
characterized by a non-Hermitian effective hamiltonian matrix which has EPs \cite{el2018non} 
when $(\gamma_2 + \gamma_1)^2/4 = |g|^2$ at eigen-frequency, $\Omega = (\omega +\epsilon +
i(\gamma_2-\gamma_1))/4$. (Note that the EP occurs at real frequencies when $\gamma_1=\gamma_2$, and
coincides with the lasing threshold). EP-enhanced sensing will be achieved by 
monitoring the quadrature amplitudes associated with $a_1,a_2$ at $\omega$ as the EP (and lasing frequency)
is shifted to the frequency $\omega + \epsilon$ via the perturbation.  Due to the EP, the resonance amplitude falls off 
as $\epsilon^{-2}$, leading to higher sensitivity.  However   Eq. \ref{eq:EOM-2} only 
describes the average behavior near the EP and not the noise properties.

The imaginary parts of the frequencies ($\gamma_1,\gamma_2$) arise from a combination
of outcoupling to the probe channels $A_{1},A_{2}$ with rate $\kappa_{1}, \kappa_{2}$, 
(for simplicity we assume $\kappa_{1}=\kappa_{2}=\kappa$) and loss and gain processes in the cavities
with rate $\eta_{1}$ ($\eta_{2}$) \footnote{We note that when $\kappa_{1}=\kappa_{2}$ the EP of the scattering
matrix is exactly the same as the EP of the non-Hermitian Hamiltonian.
However, when $\kappa_{1}\neq\kappa_{2}$, the EP of the scattering
matrix is different from the EP of the non-Hermitian Hamiltonian.
The distinction of different EPs is not the focus of this paper, and
it will be discussed in another work.}. Hence, the total loss and gain rates are $\gamma_{1}=\eta_{1}+\kappa$
and $\gamma_{2}=\eta_{2}-\kappa$ for the two modes, respectively.
We denote $A_{1,\text{in(out)}}$ and $A_{1,\text{in(out)}}$ for
the complex amplitudes of the two input (output) probe channels,
satisfying the input-output relation
\begin{equation}
\begin{pmatrix}A_{1,\text{out}}\\
A_{2,\text{out}}
\end{pmatrix}=\begin{pmatrix}A_{1,\text{in}}\\
A_{2,\text{in}}
\end{pmatrix}+\begin{pmatrix}\sqrt{\kappa}a_{1}\\
\sqrt{\kappa}a_{2}
\end{pmatrix}.\label{eq: input-output-relation}
\end{equation}
 To fully characterize the noise properties of the output, we also
need to model the fluctuations associated with the intrinsic loss
$\eta_{1}$ and gain $\eta_{2}$. As shown in Fig.~\ref{fig:1}b,
this can be done by introducing two ancillary scattering channels
$B_{1}$ and $B_{2}$ to simulate the intrinsic dissipation and amplification
processes in $a_{1}$ and $a_{2}$. According to quantum noise theory
\cite{gardiner_quantum_2004}, $B_{1}$ is coupled to $a_{1}$ via
a beam-splitter-type interaction $\hat{B}_{1}^{\dagger}\hat{a}_{1}+h.c.$
with coupling strength $\sqrt{\eta_{1}}$, while $B_{2}$ is coupled
to $a_{2}$ via a two-mode-squeezer-type interaction $\hat{B}_{2}^{\dagger}\hat{a}_{2}^{\dagger}+h.c.$
with coupling strength $\sqrt{\eta_{2}}$, and the quantum Langevin
equations for both modes can be written as
\begin{eqnarray}
\frac{d}{dt}\begin{pmatrix}\hat{a}_{1}\\
\hat{a}_{2}
\end{pmatrix} & = & -i\begin{pmatrix}\left(\omega+\epsilon\right)-i\frac{\gamma_{1}}{2} & g\\
g & \left(\omega+\epsilon\right)+i\frac{\gamma_{2}}{2}
\end{pmatrix}\begin{pmatrix}\hat{a}_{1}\\
\hat{a}_{2}
\end{pmatrix}\nonumber \\
 &  & +\begin{pmatrix}\sqrt{\kappa}\hat{A}_{1,\text{in}}\\
\sqrt{\kappa}\hat{A}_{2,\text{in}}
\end{pmatrix}+\begin{pmatrix}\sqrt{\eta_{1}}\hat{B}_{1,\text{in}}\\
-\sqrt{\eta_{2}}\hat{B}_{2,in}^{\dagger}
\end{pmatrix},\label{eq: Quantum EOM}
\end{eqnarray}
which now has additional  noise terms ($\hat{B}_{1,\text{in}}$, $\hat{B}_{2,\text{in}}^{\dagger}$)
as compared with Eq.~(\ref{eq:EOM-2}) due to noise from the reservoirs.
\paragraph{Amplitude vector and covariance matrix.}

To model the loss and gain processes, we can simply set the channels
$B_{1}$ and $B_{2}$ to zero average amplitudes, i.e. vacuum. For
any pair of bosonic annihilation operator $\hat{a}$ and creation
operator $\hat{a}^{\dagger}$, we define a pair of ``position''
quadrature operator $\hat{q}=\hat{a}+\hat{a}^{\dagger}$, and ``momentum''
quadrature operator $\hat{p}=-i\left(\hat{a}-\hat{a}^{\dagger}\right)$
\cite{weedbrook_gaussian_2012}. Given any quantum state of $N$ bosonic
modes, the amplitude vector $\mathbf{\mu}$ and covariance matrix
$\mathbf{V}$ can be defined in the quadrature basis
\begin{equation}
\mathbf{\mu}_{j}=\langle\hat{\mathbf{x}}_{j}\rangle
\end{equation}
and
\begin{equation}
\mathbf{V}_{j,k}=\frac{1}{2}\left\langle \hat{\mathbf{x}}_{j}\hat{\mathbf{x}}_{k}+\hat{\mathbf{x}}_{k}\hat{\mathbf{x}}_{j}\right\rangle -\left\langle \hat{\mathbf{x}}_{j}\right\rangle \left\langle \hat{\mathbf{x}}_{k}\right\rangle ,
\end{equation}
for $1\le j,k\le N$, with vector $\hat{\mathbf{x}}=\left(\hat{q}_{1},\hat{q}_{2},\dots,\hat{q}_{N},\hat{p}_{1},\hat{p}_{2},\dots,\hat{p}_{N}\right)^{T}$,
and $\langle\cdot\rangle$ representing the expectation value \cite{weedbrook_gaussian_2012}. 

To compute the amplitude vector and covariance matrix, we perform
the Fourier transform of Eq.~(\ref{eq: Quantum EOM}), and obtain
the relation between the Fourier transformed operators $\hat{A}\left[\omega\right]\equiv\int\hat{A}(t)e^{-i\omega t}dt$
associated with the input and output ports. The amplitude vector and
covariance matrix of the probe output channels are \cite{SM}
\begin{equation}
\mathbf{\mu}_{\text{out}}=\left(\mathbf{I}-\mathbf{G}_{\theta}\right)\mathbf{\mu}_{\text{in}}\label{eq: Amplitude}
\end{equation}
and
\begin{eqnarray}
\mathbf{V}_{\text{out}} & = & \left(\mathbf{I}-\mathbf{G}_{\theta}\right)\mathbf{V}_{\text{in}}\left(\mathbf{I}-\mathbf{G}_{\theta}\right)^{T}\nonumber \\
 &  & +\mathbf{G}_{\theta}\mathbf{R}\mathbf{V}'_{\text{in}}\mathbf{R}^{T}\mathbf{G}_{\theta}^{T},\label{eq: Covariance}
\end{eqnarray}
where $\mathbf{\mu}_{\text{in}}$ and $\mathbf{V}_{\text{in}}$ are the amplitude
vector and covariance matrix of the probe input channels ($A_{1,in}\left[\omega\right]$,
$A_{2,in}\left[\omega\right]$), $\mathbf{V}'_{\text{in}}$ is the covariance
matrix of the ancillary input channels ($B_{1}\left[\omega\right]$,
$B_{2}\left[-\omega\right]$) that induces the additional noise associated
with the loss and gain, $\mathbf{R}=\mbox{diag}\left\{ \sqrt{\eta_{1}},-\sqrt{\eta_{2}},\sqrt{\eta_{1}},\sqrt{\eta_{2}}\right\} $,
and dimensionless \emph{linear response matrix} $\mathbf{G}_{\theta}$ rescaled
by the coupling rate \cite{SM} is given by
\begin{equation}
\mathbf{G}_{\theta}=-\mathbf{\Omega}\left(\theta\mathbf{I}-\mathbf{M}\right)^{-1}
\end{equation}
with constant matrix $\mathbf{\Omega} = [[0,\mathbf{I}]; [-\mathbf{I},0]$,
the dimensionless perturbation strength $\theta=\epsilon/\kappa$,
and
\begin{equation}
\mathbf{M}=\begin{pmatrix}
0 & G & \Gamma_{1} & 0\\
G & 0 & 0 & -\Gamma_{2}\\
-\Gamma_{1} & 0 & 0 & G\\
0 & \Gamma_{2} & G & 0
\end{pmatrix}
\end{equation}
is the dimensionless effective Hamiltonian in quadrature basis with dimensionless parameters $\Gamma_{1}=\gamma_{1}/\left(2\kappa\right)$,
$\Gamma_{2}=\gamma_{2}/\left(2\kappa\right)$, and $G=g/\kappa$.

For most applications, it is sufficient to consider Gaussian states
for the probe and ancillary input channels. Since the physics process
involves only linear interactions between modes, the output states
are also Gaussian states that are completely characterized by the
amplitude vector $\mathbf{\mu}_{\text{out}}$ and covariance matrix $\mathbf{V}_{\text{out}}$
 \cite{weedbrook_gaussian_2012}. Hence, the above calculated amplitude
vector and covariance matrix are sufficient to characterize the performance
of sensing.

\paragraph{EP sensing.}

By choosing $\Gamma_{1}=\Gamma_{2}=G$ the system will be at the lasing threshold and 
will remain at an EP as the lasing frequency shifts by $\theta$ due to the perturbation 
(the situation where the system is not exactly at the lasing threshold
will be discussed later). 
Since the system remains at an EP at frequency $\omega$, we have a non-trivial Jordan decomposition
of the $\mathbf{M}$ matrix
\begin{equation}
\mathbf{M}=\mathbf{P}\begin{pmatrix}0 & 1 & 0 & 0\\
 0 & 0 & 0 & 0\\
 0 & 0  & 0 & 1\\
 0 & 0 &  0 & 0
\end{pmatrix}\mathbf{P}^{-1},
\end{equation}
with an invertible matrix $\mathbf{P}$ \footnote{For real matrix $\mathbf{M}$, there exists a real invertible matrix
$\mathbf{P}$ such that $\mathbf{P^{-1}MP}$ is a real block diagonal
matrix with each block being a real Jordan block. Such a real Jordan
form is sufficient for our investigation of EP sensing.}. 

For small perturbation $\theta\ll1$, the linear response matrix grows
as a polynomial of $\theta^{-1}$
\begin{eqnarray}
\mathbf{G}_{\theta}&=&-\mathbf{\Omega}\left(\theta\mathbf{I}-\mathbf{M}\right)^{-1}=-\theta^{-1}\sum_{n=0}^{\infty}\theta^{-n}\mathbf{\Omega}\mathbf{M}^{n} \nonumber \\
&=&-\theta^{-1}\mathbf{\Omega}-\theta^{-2}\mathbf{\Omega}\mathbf{M}, \label{eq : taylor}
\end{eqnarray}

where the second step uses the Taylor expansion and the property that
$\mathbf{M}^n=0$ for $n \ge 2$. The first term of $\mathbf{\mathbf{G}_{\theta}}$ is the general feature of a sensor at the laser threshold, while the second term implies an enhanced output signal at $\omega$ associated with the amplitude vector
{[}Eq.~(\ref{eq: Amplitude}){]} as $\theta\rightarrow0$, due to the EP.
However, the noise associated with the covariance matrix {[}Eq.~(\ref{eq: Covariance}){]}, which contains a $\theta^{-4}$ term because of its dependence on $\mathbf{G}_{\theta}$ \cite{SM}, also diverges as $\theta\rightarrow0$. Therefore, we need to 
systematically calculate the uncertainty of the measured parameter
$\theta$ in the presence of noise. In the following, we first provide a lower bound to sensitivity using
the quantum Cramer--Rao bound, and then provide an EP sensing protocol
which achieves the optimal scaling with $\theta$ as given by that quantum Cramer--Rao bound.

\paragraph{Sensitivity lower bound.}

In the presence of noise, the standard deviation
of an estimation of the parameter $\theta$, calculated from data obtained
from some measurement on a quantum state, is bounded by inverse of the
quantum Fisher information $\mathcal{I}\left(\theta\right)$ of the
state through quantum Cramer--Rao inequality \cite{braunstein_statistical_1994}
\begin{equation}
\delta\theta\ge\mathcal{I}\left(\theta\right)^{-1/2}.\label{eq: Cramer-Rao bound}
\end{equation}
For Gaussian processes, such as our scheme, the quantum Fisher information
takes the form \cite{jiang_quantum_2014,Monras13}
\[
\mathcal{I}\left(\theta\right)=\mathcal{I}_{0}\left(\theta\right)+\mathcal{I}_{1}\left(\theta\right),
\]
where $\mathcal{I}_{0}\left(\theta\right)$ is always positive and
only depends on $\mathbf{V}_{\text{out}}$, and not on $\mathbf{\mu}_{\text{out}}$,
and
\begin{equation}
\mathcal{I}_{1}\left(\theta\right)=\left(\frac{d\mathbf{\mu}_{\text{out}}}{d\theta}\right)^{T}\mathbf{V}_{\text{out}}^{-1}\frac{d\mathbf{\mu}_{\text{out}}}{d\theta}.\label{eq: Dominating term}
\end{equation}
Since $\mathcal{I}_{0}\left(\theta\right)$ only contains information
on fluctuations in the absence of the probe, the quantum Fisher information
is dominated by $\mathcal{I}_{1}\left(\theta\right)$, which can be
regarded as a generalization of the squared SNR associated with the covariance matrix.

\begin{figure}
\includegraphics[width=1\columnwidth]{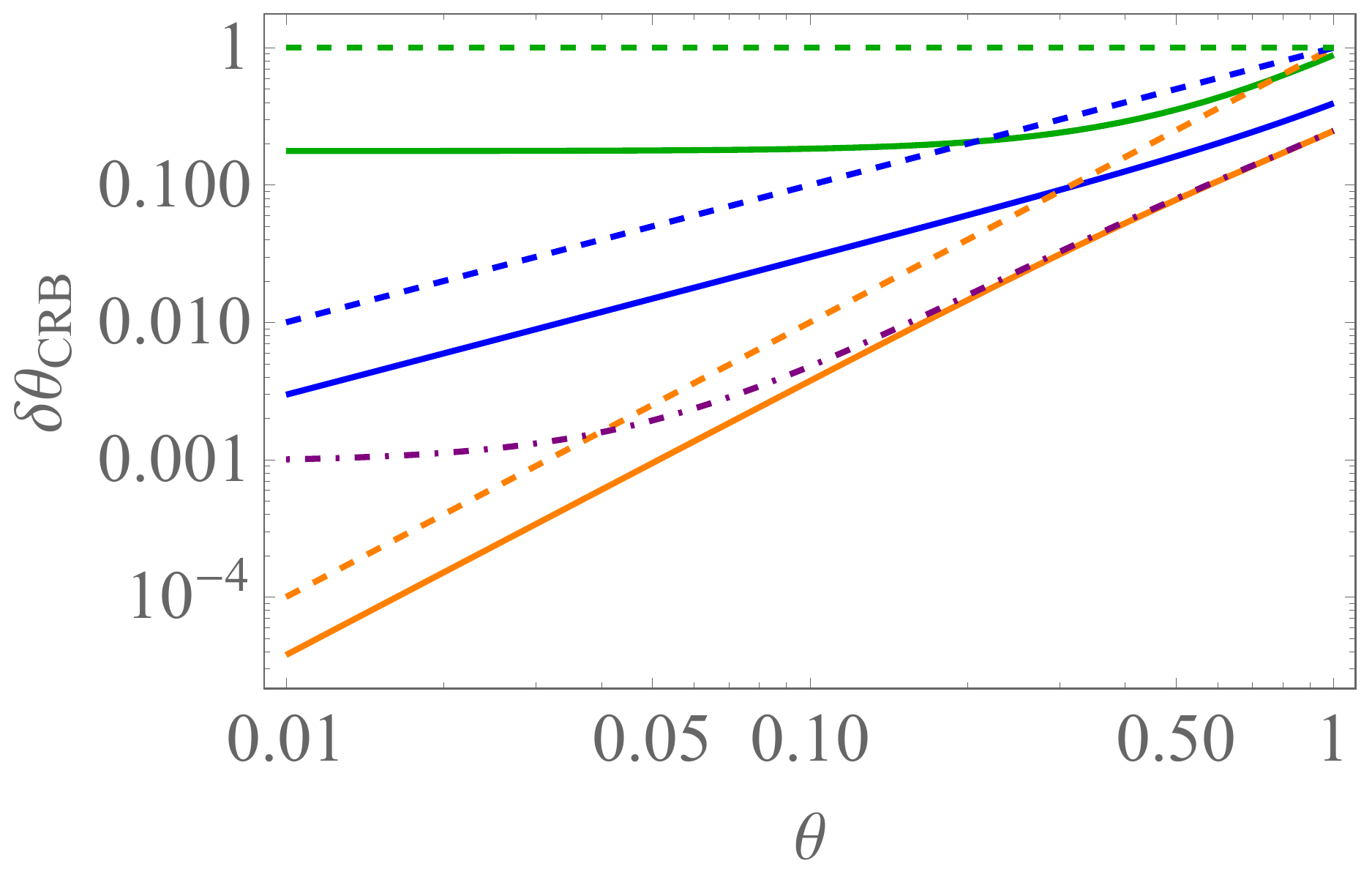}

\caption{\label{fig:2}Comparison of Cramer--Rao-bound-achieving standard variation
of parameter $\theta$ of sensing schemes. The solid orange and blue
curves are the minimal standard variation for EP sensing scheme at
the lasing threshold with all modes perturbed, EP sensing scheme at
the lasing threshold with only one mode perturbed in a two-mode system
as described in the text. The solid green curve represents a sensing
scheme without additional loss and gain in a single mode system. The
dashed orange, blue, and green lines are reference lines for $\theta^{2}$,
$\theta^{1}$, and $\theta^{0}$, which show the scaling of different
sensing schemes when the signal $\theta$ is small. The solid purple curve represents the EP sensing scheme with additional loss 
$\delta = 0.05 \Gamma$ added to each cavity, pushing the sensor below the lasing threshold.} 
\end{figure}

We plug Eqs.~(\ref{eq: Amplitude}, \ref{eq: Covariance}\&\ref{eq : taylor}) into Eq.~(\ref{eq: Dominating term})
and obtain \cite{SM}
\begin{equation}
\mathcal{I}_{1}\left(\theta\right)=\theta^{-4}\left(c_0 + O\left[\theta\right]\right),\label{eq:QFI scaling}
\end{equation}
which implies that the leading contribution to the quantum Fisher
information scales at least as $\theta^{-4}$ (orange curve in Fig.~\ref{fig:2})
and  the sensitivity, $\delta\theta\ge c_0 \times \theta^{2}$,  where the constant $c_0$ is determined by the choice of input probe signals and $c_0 > 0$  in generic situations \cite{SM}. This EP-enhanced behavior is compared with a non-EP sensing scheme
which shows no enhancement as $\theta\rightarrow0$ (green
curves in Fig.~\ref{fig:2}). Hence, EP sensing has a more favorable
lower bound than the conventional sensing protocols.

One can intuitively understand the improvement by EP sensing as a
result of correlations between different eigenmodes of the output
state. Because of the special EP structure of the linear response
matrix $\mathbf{G}_{\theta}$, different linear combinations
of the quadratures of the output state will accumulate noise with
different dependence of $\theta$. For our system, as shown in the Supplementary Material \cite{SM}, 
only two orthogonal directions in the four-dimensional input space leads to noise being
amplified $\theta^{-4}$; thus there is a large subspace of inputs 
for which the SNR is enhanced by operating at or near an EP.

\paragraph{Heterodyne detection to achieve optimized EP sensing scaling.}

The Cramer--Rao bound applies to all possible sensing schemes; now we provide a specific
EP sensing scheme that achieves the same scaling with
$\theta$ as predicted by the Cramer--Rao
bound. The idea is to use heterodyne measurement to extract the output
amplitude vector $\mu_{\text{out}}$. The covariance matrix associated with
the heterodyne detection is $\mathbf{V}_{\text{out}}+\mathbf{I}$ \cite{leonhardt1995measuring,SM},
which includes the additional quantum noise inherent in the simultaneous
measurement of both position and momentum quadratures. Fortunately,
the additional quantum noise does not depend on $\theta$, and becomes
negligibly small as compared with $\mathbf{V}_{\text{out}}$ for $\theta\rightarrow0$.
Hence, we have $\left(\mathbf{V}_{\text{out}}+\mathbf{I}\right)^{-1}=\mathbf{V}_{\text{out}}^{-1}\left(\mathbf{I}+O\left[\theta\right]\right)\approx\mathbf{V}_{\text{out}}^{-1}$.

For example, by injecting coherent state probe input with $\mu_{\text{in}}=\mathbf{P}\cdot\left(0,1,0,0\right)^{T}$,
the heterodyne detection can measure the output amplitude vector $\mu_{\text{out}}=\left( \mathbf{I}-\mathbf{G}_{\theta}\right)\mu_{\text{in}}=\mu_{\text{in}}-\mathbf{\Omega}\mathbf{P}\cdot\left(\theta^{-2},\theta^{-1},0,0\right)^{T}$.
We can obtain uncertainty $\delta\theta\approx\left[\left(\frac{d\mathbf{\mu}_{\text{out}}}{d\theta}\right)^{T}\mathbf{V}_{\text{out}}^{-1}\frac{d\mathbf{\mu}_{\text{out}}}{d\theta}\right]^{-1/2}\sim\text{\ensuremath{\theta}}^{2}$,
which has the same scaling as the lower bound obtained from Eqs.~(\ref{eq: Cramer-Rao bound},\ref{eq: Dominating term},\ref{eq:QFI scaling}).

\paragraph{General approach and higher-order EP sensing.}

We summarize our general approach to achieve EP sensing with the scaling
as obtained from the Cramer--Rao bound. For a given EP sensing scheme
based on a Gaussian process, we calculate the corresponding matrices
that can help us track the change of amplitude and covariance matrix.
Then we can calculate the quantum Fisher information and obtain the
precision bound. For a general EP sensing scheme, $\mathbf{G}_{\theta}=-\mathbf{\Omega}\left(\theta \mathbf{\Pi}-\mathbf{M}\right)^{-1}$
(in our previous discussion, $\mathbf{\Pi}=\mathbf{I}$, but in general
it can be any invertible matrix), and $\mathbf{M}\mathbf{\Pi}^{-1}=\mathbf{P}\Lambda\mathbf{P}^{-1}$,
with $\mathbf{P}$ invertible , and $\mathbf{\Lambda}$ known as
the Jordan normal form of $\mathbf{M}\mathbf{\Pi}^{-1}$ consisting of diagonal blocks of size $N_{i}$ (for the $i$th block),
each with eigenvalue zero. When $N_{i}=1$, the corresponding block
is just a scalar, which is not an EP. To have EP enhanced sensing,
we need at least one non-trivial Jordan block ($N_{i}\ge2)$ with
eigenvalue zero. 

Let $N=\text{max}_{i}N_{i}$ be the size of the largest zero-eigenvalue
Jordan block, which corresponds to the $\left(N-1\right)$-th order
EP. Then it is easy to show that $\mathbf{G}_{\theta}=\theta^{-N}\left(-\mathbf{\Omega}\mathbf{C}_{0}+O\left(\theta\right)\right)+\cdots$
with $\theta\rightarrow0$ and $\mathbf{C}_{0}$ a constant matrix.
This divergence near $\theta=0$ leads to $\theta^{-N}$ amplification
of the amplitude and $\theta^{-2N}$ amplification of the covariance
matrix. One might be tempted to argue that the $\mathcal{I}_{1}\left(\theta\right)$
is then proportional to $\theta^{-2}$ since the scaling of amplification
with $N$ can be perfectly canceled by covariance matrix. However,
a more rigorous calculation shows this is over pessimistic. As $d\mathbf{G}_{\theta}/d\theta=\mathbf{G}_{\theta}\mathbf{\Pi}\mathbf{\Omega}\mathbf{G}_{\theta}$,
only one of the $\mathbf{G}_{\theta}$ cancel with the amplification
in the covariance matrix. We have $\mathcal{I}_{1}\left(\theta\right)=\mathbf{\mu}_{\text{in}}^{T}\mathbf{G}_{\theta}^{T}\mathbf{C}_{1}\mathbf{G}_{\theta}\mathbf{\mu}_{\text{in}}$,
where $\mathbf{C}_{1}$ is a positive definite matrix \cite{SM}.
So one can conclude that 
\begin{equation}
\mathcal{I}_{1}\left(\theta\right)\approx\theta^{-2N}
\end{equation}
for $\left(N-1\right)$-th order EP. Since $\mathcal{I}_{0}\left(\theta\right)$
is always positive, $\mathcal{I}_{1}\left(\theta\right)$ gives a
lower bound on quantum Fisher information. We have then the quantum
Cramer--Rao bound $\delta\theta\gtrsim\theta^{N}$ \footnote{Actually, we can show that the ``\textasciitilde{}'' sign holds
if the process is very noisy \cite{SM}}, the scaling of which can be achieved by performing heterodyne measurement
on all of the outputs even in this general situation.

\paragraph{Robustness of enhanced sensing}

The main assumptions used to realize enhanced sensing is precise tuning to an EP
and operating the sensor at the lasing threshold.  As noted above, if this is realized, there is a large family of input states which
generate enhanced sensitivity for appropriately chose outputs, so the sensor will be robust to the input state used.
While operation at the lasing threshold condition is assumed in order to derive the
results above, we can actually relax this condition, and still achieve sensitivity enhancement over some parameter range.
A small detuning $\delta$ from the lasing threshold will simply cut off the enhanced sensing for small $\theta$.
As shown in Fig.~\ref{fig:2}, (purple curve) when we introduce additional loss $\delta$ to both cavities, the quantum
Fisher information is upper bounded by $\mathcal{I}^{UB}\approx\left\lVert \mathbf{G}_{\theta=0}\right \rVert^2\approx\delta^{-4}$, where $\lVert\cdot \rVert$ represents the trace norm\cite{SM}. When the system is sufficiently close to the lasing threshold
($\delta\ll\theta$), the quantum Fisher information does not exceed
the upper bound ($\mathcal{I}\approx\theta^{-4}\ll\mathcal{I}^{UB}$),
so that we will have EP enhanced sensing. Generally, for higher order
EPs, we will have EP enhanced sensing near the lasing threshold as
long as $\left\lVert\mathbf{G}_{\theta}\right\rVert\gg\theta^{-N}$
\cite{SM}. 

In conclusion, we have established a theoretical framework using
quantum noise theory to calculate systematically both the signal and noise
of EP sensors, operating near the lasing threshold. Using the quantum Fisher information, we have obtained
the lower bound of ultimate sensitivity of EP-sensors. Moreover,
we provide a heterodyne detection scheme to achieve the optimal scaling
of the sensitivity predicted by this bound. Since these EP sensors are described by Gaussian
processes with linear interactions, EP sensing near the laser
threshold coupled with heterodyne detection should be feasible with
the current experimental techniques.

Note added: During the completion of this work, we became aware of
a related study by Lau and Clerk \cite{lau2018fundamental}.  

\begin{acknowledgments}
We would like to thank Aash Clerk, Hoi-Kwan Lau, Jack Harris, Jianming
Wen, and Peter Rabl for discussions. We also acknowledge support from
the ARL-CDQI, ARO (W911NF-14-1-0011, W911NF-16-1-0563), AFOSR MURI
(FA9550-14-1-0052, FA9550-15-1-0015), ARO MURI (W911NF-16-1-0349),
NSF (EFMA-1640959), Alfred P. Sloan Foundation (BR2013-0049), and
Packard Foundation (2013-39273).
\end{acknowledgments}

\bibliographystyle{apsrev}
\bibliography{MyLibrary}

\end{document}